\documentclass[12pt]{article}

\usepackage[safe]{tipa}
\usepackage{amstext}
\usepackage{times}
\usepackage{helvet}
\usepackage{courier}
\usepackage{amsmath, amssymb}
\usepackage{latexsym}
\usepackage{eurosym}
\usepackage{wasysym}
\usepackage{multirow}

\makeatletter
\newif\if@restonecol
\makeatother

\usepackage[ruled,lined,boxed]{algorithm2e}
\usepackage{tikz}
\usepackage{pslatex}
\usepackage[arrow, matrix, curve]{xy}
\usepackage{epsfig}
\usepackage[pdfmark]{thumbpdf}
\usepackage[lflt]{floatflt}

\title{Towards Barter Double Auction as Model for Bilateral Social Cooperations}
\author{Rustam Tagiew\\
Institute for Computer Science, TU Bergakademie Freiberg,\\
tagiew@informatik.tu-freiberg.de
}
\date{\today}

\begin{document}
\maketitle

\begin{abstract}
The idea of this paper is an advanced game concept. This concept is expected to model non-monetary bilateral cooperations between self-interested agents. Such non-monetary cases are social cooperations like allocation of high level jobs or sexual relationships among humans. In a barter double auction, there is a big amount of agents. Every agent has a vector of parameters which specifies his demand and a vector which specifies his offer. Two agents can achieve a commitment through barter exchange. The subjective satisfaction level (a number between 0\% and 100\%) of an agent is as high as small is the distance between his demand and the accepted offer. This paper introduces some facets of this complex game concept.
\end{abstract}
\section{Introduction}
\indent In the artificial intelligence, an agent can be designed to be humanlike or rational \cite[p.2]{russel}. Rationality means that an agent maximizes his payoff or the chance of achieving his goal considering what he knows and perceives. Rational and humanlike is by far not the same. It is this (seeming) absence of rationality in human behavior. ''British people argue that it is worth spending billions of pounds to improve the safety of the rail system. However, the same people habitually travel by car rather than by train, even though travelling by car is approximately 30 times more dangerous than by train!''\cite[p.527--530]{hrational} Nevertheless, economists model business interactions as interactions between rational self-interested agents. For instance, a loyal manager is expected to choose the action which maximizes the profit of his firm.\\
\indent Achieving a goal makes an agent 'happy'. ''Because 'happy' does not sound very scientific, the customary terminology is to say that if one world state is preferred to another, then it has higher utility for the agent'' \cite[p.51]{russel}. For (small group) strategic interactions, economists use the game theory. In the game theory the utility is also called payoff. The game theory models rational behavior in strategic interactions. One can say that an agent will be twice as 'happy', if he gets a doubled payoff. The experimenters which try to explain the difference between the human behavior and the game theoretic predictions accept the previous sentence at face value and pay their subjects according to their performance \cite{vspt}. But the money is not a measure for the happiness \cite{money}. The safety is also not a messure for the happiness, as it has been mentioned in the rail system example. Payoff is not bounded, happiness is - you can be happy or unhappy. For non-monetary interactions, happiness can be quantified by a satisfaction level which is a number between 0\% and 100\% \cite{happiness}.\\  
\indent This work addresses the question how to reason strategically if one wants to maximize his level of satisfaction in a non-monetary domain. Strategic reasoning means taking in account the reasoning of the other agents \cite{rubin}. Participants of a non-monetary strategic interaction are interested in achieving the highest satisfaction level and their success is in part depending on decisions of other participants. This issues can be studied by game theory. One can say without any doubt that if a human player is trained in a concrete game, he performs close to optimal. And ''... when the chips are down, the payoff is not five dollars but a successful career, and people have time to understand the situation-the predictions of game theory fare quite well.''[Robert Aumann]. Bilateral cooperations are chosen as a concrete example for a non-monetary strategic interaction. How can an agent get as much 'happy' as possible in a non-monetary bilateral interaction? For answering this question a new game concept is introduced - the barter double auction. It is not save that humans behave fully rationally in the barter double auction. Hence, this paper discusses the human behavior in an intuitive semi-formal way.  
\paragraph{Outline}
\indent The chapter 'previous work' is skipped, because there no similar concepts found at time. The next chapter introduces the mathematical concept of the barter double auction. The chapter \ref{casestudy} contains a preliminary summary of extreme cases for barter double auction and explains possible solutions for them. In chapter \ref{conclusions}, there is set of conclusions for this work.
\section{Barter Double Auction}\label{bda}
\indent There is a couple of agents. Let $N$ be the number of agents. Each agent $i$ in this set has an offer $O_i$ and a demand $D_i$. The offer is defined as a vector in a multi dimensional space. Every dimesion embodies one certain characteristic of the offer. The demand is defined in the same way. If two agents $i$ and $j$ cooperate, the agent $i$ gets the offer $O_j$ and the agent $j$ gets the offer $O_i$. The cooperation is a barter exchange. If an agent $i$ gets an offer which is equal to his demand, he gets what he needs. This means that his satisfaction level $S_i$ is 100\%. But what if not? Using the Euclidean distance $\text{dist}(D_i,O_j)$ between the demand $D_i$ and the offer $O_j$, one can approximate mathematically the relation between the satisfaction level and the dismatch of the offer as following:\\ 
\[
 S_i = e^{-{\alpha_i}*{\text{dist}(D_i,O_j)}^2}
\]
\begin{figure}
\begin{center}
\resizebox{210pt}{!}{\input{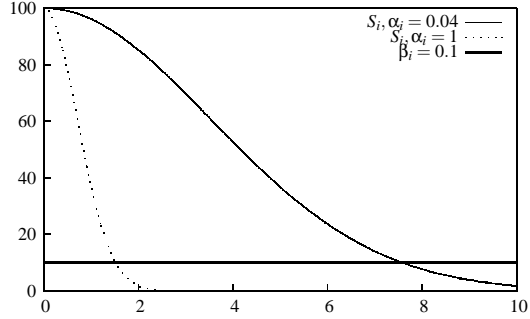}}
\caption[Satisfaction level.]{Satisfaction level.}
\label{satisfac}
\end{center}
\end{figure}
\begin{table}
\begin{center}
\begin{tabular}{|r|lr|lr|}
  \hline
Alice$\setminus$Bob &   \multicolumn{2}{c}{Allure/Accept} \vline &\multicolumn{2}{c}{Ignore}\vline \\
  \hline
\multirow{2}{30mm}{\hspace{\stretch{1} Allure/Accept}}  &   & $S_{Bob}$&  & $\beta_{Bob}$ \\
                                              &  $S_{Alice}$ &  &$\beta_{Alice}*\gamma$ &  \\
  \hline
\multirow{2}{30mm}{\hspace{\stretch{1}}Ignore}  &   & $\beta_{Bob}*\gamma$ &  & $\beta_{Bob}$\\
                                              & $\beta_{Alice}$ &  & $\beta_{Alice}$ &  \\
  \hline
\end{tabular}
\caption[Satisfaction matrix.]{Satisfaction matrix for a bilateral non-monetary cooperation.}
\label{matrix}
\end{center}
\end{table}
\indent $\alpha_i>0$ is a constant value which defines the sensitivity of the agent's satisfaction level to the dismatch of the offer. $0<\beta_i<1$ will be the default satisfaction level, if the agent does not cooperate. Fig.\ref{satisfac} shows the plots of the satisfaction level for different $\alpha$. It shows also the plot of $\beta$. As you see, if the dismatch of the offer achieves a certain level, an agent will not accept it. Two agents Alice and Bob will cooperate assuming rationality, if the satisfaction levels of both agents exceed the noncooperative case. Tab.\ref{matrix} shows the satisfaction matrix. The lower left corners of this matrix contain outcomes of Alice and the upper right corners contain the outcomes of Bob. Alice can choose between rows and Bob can choose between columns. Each agent can eather allure(accept) or ignore a cooperation with the other agent. Alluring an ignoring agent diminishes the non-cooperative satisfaction level. This value $0<\gamma_i<1$ is called the frustration factor of an agent. Game theoretically considered, there are two equilibria in pure strategies at most - mutual alluring(accepting) or ignoring. If an agent $i$ allures $m$ different agents and gets no results, he achieves the satisfaction level ${\beta_i}*{\gamma_i}^m$.\\
\indent In the double auction case, every agent allures agents with the best matching offer and is allured on the other side by other agents. The protocol for an agent in the barter double auction is following: 
\begin{enumerate}
  \item Allure a subset of agents.
  \item Accept one or none of the alluring agents. Ignore the rest.
  \item Confirm one of the accepts. Defect the rest.
\end{enumerate}
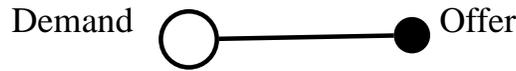
\begin{figure}
\begin{center}
\resizebox{210pt}{!}{
\ifx\du\undefined
  \newlength{\du}
\fi
\setlength{\du}{15\unitlength}
\begin{tikzpicture}
\pgftransformxscale{1.000000}
\pgftransformyscale{-1.000000}
\definecolor{dialinecolor}{rgb}{0.000000, 0.000000, 0.000000}
\pgfsetstrokecolor{dialinecolor}
\definecolor{dialinecolor}{rgb}{1.000000, 1.000000, 1.000000}
\pgfsetfillcolor{dialinecolor}
\pgfsetlinewidth{0.100000\du}
\pgfsetdash{}{0pt}
\pgfsetdash{}{0pt}
\pgfsetbuttcap
\pgfsetmiterjoin
\pgfsetlinewidth{0.100000\du}
\pgfsetbuttcap
\pgfsetmiterjoin
\pgfsetdash{}{0pt}
\definecolor{dialinecolor}{rgb}{1.000000, 1.000000, 1.000000}
\pgfsetfillcolor{dialinecolor}
\pgfpathellipse{\pgfpoint{10.575000\du}{6.275000\du}}{\pgfpoint{0.625000\du}{0\du}}{\pgfpoint{0\du}{0.625000\du}}
\pgfusepath{fill}
\definecolor{dialinecolor}{rgb}{0.000000, 0.000000, 0.000000}
\pgfsetstrokecolor{dialinecolor}
\pgfpathellipse{\pgfpoint{10.575000\du}{6.275000\du}}{\pgfpoint{0.625000\du}{0\du}}{\pgfpoint{0\du}{0.625000\du}}
\pgfusepath{stroke}
\pgfsetlinewidth{0.010000\du}
\pgfsetbuttcap
\pgfsetmiterjoin
\pgfsetdash{}{0pt}
\definecolor{dialinecolor}{rgb}{0.000000, 0.000000, 0.000000}
\pgfsetstrokecolor{dialinecolor}
\pgfpathellipse{\pgfpoint{10.575000\du}{6.275000\du}}{\pgfpoint{0.625000\du}{0\du}}{\pgfpoint{0\du}{0.625000\du}}
\pgfusepath{stroke}
\pgfsetlinewidth{0.100000\du}
\pgfsetdash{}{0pt}
\pgfsetdash{}{0pt}
\pgfsetbuttcap
{
\definecolor{dialinecolor}{rgb}{0.000000, 0.000000, 0.000000}
\pgfsetfillcolor{dialinecolor}
\definecolor{dialinecolor}{rgb}{0.000000, 0.000000, 0.000000}
\pgfsetstrokecolor{dialinecolor}
\draw (11.249300\du,6.262310\du)--(15.151300\du,6.188870\du);
}
\pgfsetlinewidth{0.100000\du}
\pgfsetdash{}{0pt}
\pgfsetdash{}{0pt}
\pgfsetbuttcap
\pgfsetmiterjoin
\pgfsetlinewidth{0.100000\du}
\pgfsetbuttcap
\pgfsetmiterjoin
\pgfsetdash{}{0pt}
\definecolor{dialinecolor}{rgb}{0.000000, 0.000000, 0.000000}
\pgfsetfillcolor{dialinecolor}
\pgfpathellipse{\pgfpoint{15.556250\du}{6.181250\du}}{\pgfpoint{0.356250\du}{0\du}}{\pgfpoint{0\du}{0.356250\du}}
\pgfusepath{fill}
\definecolor{dialinecolor}{rgb}{0.000000, 0.000000, 0.000000}
\pgfsetstrokecolor{dialinecolor}
\pgfpathellipse{\pgfpoint{15.556250\du}{6.181250\du}}{\pgfpoint{0.356250\du}{0\du}}{\pgfpoint{0\du}{0.356250\du}}
\pgfusepath{stroke}
\pgfsetlinewidth{0.010000\du}
\pgfsetbuttcap
\pgfsetmiterjoin
\pgfsetdash{}{0pt}
\definecolor{dialinecolor}{rgb}{0.000000, 0.000000, 0.000000}
\pgfsetstrokecolor{dialinecolor}
\pgfpathellipse{\pgfpoint{15.556250\du}{6.181250\du}}{\pgfpoint{0.356250\du}{0\du}}{\pgfpoint{0\du}{0.356250\du}}
\pgfusepath{stroke}
\definecolor{dialinecolor}{rgb}{0.000000, 0.000000, 0.000000}
\pgfsetstrokecolor{dialinecolor}
\node[anchor=west] at (15.900000\du,5.850000\du){Offer};
\definecolor{dialinecolor}{rgb}{0.000000, 0.000000, 0.000000}
\pgfsetstrokecolor{dialinecolor}
\node[anchor=west] at (6.350000\du,5.850000\du){Demand};
\end{tikzpicture}}
\caption[Graphical representation of an agent's type.]{Graphical representation of an agent's type.}
\label{grarep}
\end{center}
\end{figure}
\indent To understand the barter double auction, one needs a graphical representation. The number of dimensions is reduced to two. A demand is a circle and an offer is a point on a surface. If a couple of agents have a certain demand and a certain offer, they will be represented through a connection between a circle and a point. Fig.\ref{grarep} shows a type of a couple of agents.\\
\section{Case Study}\label{casestudy}
\indent As first, let us consider the simplest example of the barter double auction. A couple of children want to play with seesaws. For playing with a seesaw, you need a partner which has approximately the same weight. The partner is interested for the same. In this scenario, the demand of an agent is equal to his offer. Fig.\ref{seesaw} shows this type. If the children are all of the same weight, they will have to coordinate. Due to the fact that there is no difference between partners, a agent would randomly choose a partner. In this case, every agent allures only one agent and he will allure none, if he is allready allured. If the number of agents is not even, then there will be one agent which gets nobody for a cooperation. If there are many children with different weights, every agent will cooperate with the best matching agent. Fig.\ref{setseesaw} shows such an example. If there is only one agent for each type, there will be two agents (of type 1 and 4) which will get a very low satisfaction level, if they cooperate.\\  
\begin{figure}
\begin{center}
\resizebox{40pt}{!}{
\ifx\du\undefined
  \newlength{\du}
\fi
\setlength{\du}{15\unitlength}
\begin{tikzpicture}
\pgftransformxscale{1.000000}
\pgftransformyscale{-1.000000}
\definecolor{dialinecolor}{rgb}{0.000000, 0.000000, 0.000000}
\pgfsetstrokecolor{dialinecolor}
\definecolor{dialinecolor}{rgb}{1.000000, 1.000000, 1.000000}
\pgfsetfillcolor{dialinecolor}
\pgfsetlinewidth{0.100000\du}
\pgfsetdash{}{0pt}
\pgfsetdash{}{0pt}
\pgfsetbuttcap
\pgfsetmiterjoin
\pgfsetlinewidth{0.100000\du}
\pgfsetbuttcap
\pgfsetmiterjoin
\pgfsetdash{}{0pt}
\definecolor{dialinecolor}{rgb}{1.000000, 1.000000, 1.000000}
\pgfsetfillcolor{dialinecolor}
\pgfpathellipse{\pgfpoint{16.537500\du}{6.012500\du}}{\pgfpoint{1.487500\du}{0\du}}{\pgfpoint{0\du}{1.487500\du}}
\pgfusepath{fill}
\definecolor{dialinecolor}{rgb}{0.000000, 0.000000, 0.000000}
\pgfsetstrokecolor{dialinecolor}
\pgfpathellipse{\pgfpoint{16.537500\du}{6.012500\du}}{\pgfpoint{1.487500\du}{0\du}}{\pgfpoint{0\du}{1.487500\du}}
\pgfusepath{stroke}
\pgfsetlinewidth{0.010000\du}
\pgfsetbuttcap
\pgfsetmiterjoin
\pgfsetdash{}{0pt}
\definecolor{dialinecolor}{rgb}{0.000000, 0.000000, 0.000000}
\pgfsetstrokecolor{dialinecolor}
\pgfpathellipse{\pgfpoint{16.537500\du}{6.012500\du}}{\pgfpoint{1.487500\du}{0\du}}{\pgfpoint{0\du}{1.487500\du}}
\pgfusepath{stroke}
\pgfsetlinewidth{0.100000\du}
\pgfsetdash{}{0pt}
\pgfsetdash{}{0pt}
\pgfsetbuttcap
\pgfsetmiterjoin
\pgfsetlinewidth{0.100000\du}
\pgfsetbuttcap
\pgfsetmiterjoin
\pgfsetdash{}{0pt}
\definecolor{dialinecolor}{rgb}{0.000000, 0.000000, 0.000000}
\pgfsetfillcolor{dialinecolor}
\pgfpathellipse{\pgfpoint{16.537500\du}{6.062500\du}}{\pgfpoint{0.737500\du}{0\du}}{\pgfpoint{0\du}{0.737500\du}}
\pgfusepath{fill}
\definecolor{dialinecolor}{rgb}{0.000000, 0.000000, 0.000000}
\pgfsetstrokecolor{dialinecolor}
\pgfpathellipse{\pgfpoint{16.537500\du}{6.062500\du}}{\pgfpoint{0.737500\du}{0\du}}{\pgfpoint{0\du}{0.737500\du}}
\pgfusepath{stroke}
\pgfsetlinewidth{0.010000\du}
\pgfsetbuttcap
\pgfsetmiterjoin
\pgfsetdash{}{0pt}
\definecolor{dialinecolor}{rgb}{0.000000, 0.000000, 0.000000}
\pgfsetstrokecolor{dialinecolor}
\pgfpathellipse{\pgfpoint{16.537500\du}{6.062500\du}}{\pgfpoint{0.737500\du}{0\du}}{\pgfpoint{0\du}{0.737500\du}}
\pgfusepath{stroke}
\pgfsetlinewidth{0.100000\du}
\pgfsetdash{}{0pt}
\pgfsetdash{}{0pt}
\pgfsetbuttcap
{
\definecolor{dialinecolor}{rgb}{0.000000, 0.000000, 0.000000}
\pgfsetfillcolor{dialinecolor}
\definecolor{dialinecolor}{rgb}{0.000000, 0.000000, 0.000000}
\pgfsetstrokecolor{dialinecolor}
\draw (18.025000\du,6.012500\du)--(17.323376\du,6.036084\du);
}
\end{tikzpicture}}
\caption[Seesaw type.]{Seesaw type.}
\label{seesaw}
\end{center}
\end{figure}
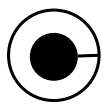
\begin{figure}
\begin{center}
\resizebox{180pt}{!}{
\ifx\du\undefined
  \newlength{\du}
\fi
\setlength{\du}{15\unitlength}
\begin{tikzpicture}
\pgftransformxscale{1.000000}
\pgftransformyscale{-1.000000}
\definecolor{dialinecolor}{rgb}{0.000000, 0.000000, 0.000000}
\pgfsetstrokecolor{dialinecolor}
\definecolor{dialinecolor}{rgb}{1.000000, 1.000000, 1.000000}
\pgfsetfillcolor{dialinecolor}
\pgfsetlinewidth{0.100000\du}
\pgfsetdash{}{0pt}
\pgfsetdash{}{0pt}
\pgfsetbuttcap
\pgfsetmiterjoin
\pgfsetlinewidth{0.100000\du}
\pgfsetbuttcap
\pgfsetmiterjoin
\pgfsetdash{}{0pt}
\definecolor{dialinecolor}{rgb}{0.000000, 0.000000, 0.000000}
\pgfsetstrokecolor{dialinecolor}
\pgfpathellipse{\pgfpoint{9.172500\du}{5.982500\du}}{\pgfpoint{1.487500\du}{0\du}}{\pgfpoint{0\du}{1.487500\du}}
\pgfusepath{stroke}
\pgfsetlinewidth{0.010000\du}
\pgfsetbuttcap
\pgfsetmiterjoin
\pgfsetdash{}{0pt}
\definecolor{dialinecolor}{rgb}{0.000000, 0.000000, 0.000000}
\pgfsetstrokecolor{dialinecolor}
\pgfpathellipse{\pgfpoint{9.172500\du}{5.982500\du}}{\pgfpoint{1.487500\du}{0\du}}{\pgfpoint{0\du}{1.487500\du}}
\pgfusepath{stroke}
\pgfsetlinewidth{0.100000\du}
\pgfsetdash{}{0pt}
\pgfsetdash{}{0pt}
\pgfsetbuttcap
\pgfsetmiterjoin
\pgfsetlinewidth{0.100000\du}
\pgfsetbuttcap
\pgfsetmiterjoin
\pgfsetdash{}{0pt}
\definecolor{dialinecolor}{rgb}{0.000000, 0.000000, 0.000000}
\pgfsetfillcolor{dialinecolor}
\pgfpathellipse{\pgfpoint{9.172500\du}{6.032500\du}}{\pgfpoint{0.737500\du}{0\du}}{\pgfpoint{0\du}{0.737500\du}}
\pgfusepath{fill}
\definecolor{dialinecolor}{rgb}{0.000000, 0.000000, 0.000000}
\pgfsetstrokecolor{dialinecolor}
\pgfpathellipse{\pgfpoint{9.172500\du}{6.032500\du}}{\pgfpoint{0.737500\du}{0\du}}{\pgfpoint{0\du}{0.737500\du}}
\pgfusepath{stroke}
\pgfsetlinewidth{0.010000\du}
\pgfsetbuttcap
\pgfsetmiterjoin
\pgfsetdash{}{0pt}
\definecolor{dialinecolor}{rgb}{0.000000, 0.000000, 0.000000}
\pgfsetstrokecolor{dialinecolor}
\pgfpathellipse{\pgfpoint{9.172500\du}{6.032500\du}}{\pgfpoint{0.737500\du}{0\du}}{\pgfpoint{0\du}{0.737500\du}}
\pgfusepath{stroke}
\pgfsetlinewidth{0.100000\du}
\pgfsetdash{}{0pt}
\pgfsetdash{}{0pt}
\pgfsetbuttcap
{
\definecolor{dialinecolor}{rgb}{0.000000, 0.000000, 0.000000}
\pgfsetfillcolor{dialinecolor}
\definecolor{dialinecolor}{rgb}{0.000000, 0.000000, 0.000000}
\pgfsetstrokecolor{dialinecolor}
\draw (10.660000\du,5.982500\du)--(9.958376\du,6.006084\du);
}
\pgfsetlinewidth{0.100000\du}
\pgfsetdash{}{0pt}
\pgfsetdash{}{0pt}
\pgfsetbuttcap
\pgfsetmiterjoin
\pgfsetlinewidth{0.100000\du}
\pgfsetbuttcap
\pgfsetmiterjoin
\pgfsetdash{}{0pt}
\definecolor{dialinecolor}{rgb}{0.000000, 0.000000, 0.000000}
\pgfsetstrokecolor{dialinecolor}
\pgfpathellipse{\pgfpoint{26.412500\du}{6.147500\du}}{\pgfpoint{1.487500\du}{0\du}}{\pgfpoint{0\du}{1.487500\du}}
\pgfusepath{stroke}
\pgfsetlinewidth{0.010000\du}
\pgfsetbuttcap
\pgfsetmiterjoin
\pgfsetdash{}{0pt}
\definecolor{dialinecolor}{rgb}{0.000000, 0.000000, 0.000000}
\pgfsetstrokecolor{dialinecolor}
\pgfpathellipse{\pgfpoint{26.412500\du}{6.147500\du}}{\pgfpoint{1.487500\du}{0\du}}{\pgfpoint{0\du}{1.487500\du}}
\pgfusepath{stroke}
\pgfsetlinewidth{0.100000\du}
\pgfsetdash{}{0pt}
\pgfsetdash{}{0pt}
\pgfsetbuttcap
\pgfsetmiterjoin
\pgfsetlinewidth{0.100000\du}
\pgfsetbuttcap
\pgfsetmiterjoin
\pgfsetdash{}{0pt}
\definecolor{dialinecolor}{rgb}{0.000000, 0.000000, 0.000000}
\pgfsetfillcolor{dialinecolor}
\pgfpathellipse{\pgfpoint{26.427500\du}{6.182500\du}}{\pgfpoint{0.752500\du}{0\du}}{\pgfpoint{0\du}{0.752500\du}}
\pgfusepath{fill}
\definecolor{dialinecolor}{rgb}{0.000000, 0.000000, 0.000000}
\pgfsetstrokecolor{dialinecolor}
\pgfpathellipse{\pgfpoint{26.427500\du}{6.182500\du}}{\pgfpoint{0.752500\du}{0\du}}{\pgfpoint{0\du}{0.752500\du}}
\pgfusepath{stroke}
\pgfsetlinewidth{0.010000\du}
\pgfsetbuttcap
\pgfsetmiterjoin
\pgfsetdash{}{0pt}
\definecolor{dialinecolor}{rgb}{0.000000, 0.000000, 0.000000}
\pgfsetstrokecolor{dialinecolor}
\pgfpathellipse{\pgfpoint{26.427500\du}{6.182500\du}}{\pgfpoint{0.752500\du}{0\du}}{\pgfpoint{0\du}{0.752500\du}}
\pgfusepath{stroke}
\pgfsetlinewidth{0.100000\du}
\pgfsetdash{}{0pt}
\pgfsetdash{}{0pt}
\pgfsetbuttcap
{
\definecolor{dialinecolor}{rgb}{0.000000, 0.000000, 0.000000}
\pgfsetfillcolor{dialinecolor}
\definecolor{dialinecolor}{rgb}{0.000000, 0.000000, 0.000000}
\pgfsetstrokecolor{dialinecolor}
\draw (27.900000\du,6.147500\du)--(27.228459\du,6.163462\du);
}
\pgfsetlinewidth{0.100000\du}
\pgfsetdash{}{0pt}
\pgfsetdash{}{0pt}
\pgfsetbuttcap
\pgfsetmiterjoin
\pgfsetlinewidth{0.100000\du}
\pgfsetbuttcap
\pgfsetmiterjoin
\pgfsetdash{}{0pt}
\definecolor{dialinecolor}{rgb}{0.000000, 0.000000, 0.000000}
\pgfsetstrokecolor{dialinecolor}
\pgfpathellipse{\pgfpoint{16.272500\du}{6.072500\du}}{\pgfpoint{1.487500\du}{0\du}}{\pgfpoint{0\du}{1.487500\du}}
\pgfusepath{stroke}
\pgfsetlinewidth{0.010000\du}
\pgfsetbuttcap
\pgfsetmiterjoin
\pgfsetdash{}{0pt}
\definecolor{dialinecolor}{rgb}{0.000000, 0.000000, 0.000000}
\pgfsetstrokecolor{dialinecolor}
\pgfpathellipse{\pgfpoint{16.272500\du}{6.072500\du}}{\pgfpoint{1.487500\du}{0\du}}{\pgfpoint{0\du}{1.487500\du}}
\pgfusepath{stroke}
\pgfsetlinewidth{0.100000\du}
\pgfsetdash{}{0pt}
\pgfsetdash{}{0pt}
\pgfsetbuttcap
\pgfsetmiterjoin
\pgfsetlinewidth{0.100000\du}
\pgfsetbuttcap
\pgfsetmiterjoin
\pgfsetdash{}{0pt}
\definecolor{dialinecolor}{rgb}{0.000000, 0.000000, 0.000000}
\pgfsetfillcolor{dialinecolor}
\pgfpathellipse{\pgfpoint{16.272500\du}{6.122500\du}}{\pgfpoint{0.737500\du}{0\du}}{\pgfpoint{0\du}{0.737500\du}}
\pgfusepath{fill}
\definecolor{dialinecolor}{rgb}{0.000000, 0.000000, 0.000000}
\pgfsetstrokecolor{dialinecolor}
\pgfpathellipse{\pgfpoint{16.272500\du}{6.122500\du}}{\pgfpoint{0.737500\du}{0\du}}{\pgfpoint{0\du}{0.737500\du}}
\pgfusepath{stroke}
\pgfsetlinewidth{0.010000\du}
\pgfsetbuttcap
\pgfsetmiterjoin
\pgfsetdash{}{0pt}
\definecolor{dialinecolor}{rgb}{0.000000, 0.000000, 0.000000}
\pgfsetstrokecolor{dialinecolor}
\pgfpathellipse{\pgfpoint{16.272500\du}{6.122500\du}}{\pgfpoint{0.737500\du}{0\du}}{\pgfpoint{0\du}{0.737500\du}}
\pgfusepath{stroke}
\pgfsetlinewidth{0.100000\du}
\pgfsetdash{}{0pt}
\pgfsetdash{}{0pt}
\pgfsetbuttcap
{
\definecolor{dialinecolor}{rgb}{0.000000, 0.000000, 0.000000}
\pgfsetfillcolor{dialinecolor}
\definecolor{dialinecolor}{rgb}{0.000000, 0.000000, 0.000000}
\pgfsetstrokecolor{dialinecolor}
\draw (17.760000\du,6.072500\du)--(17.058376\du,6.096084\du);
}
\pgfsetlinewidth{0.100000\du}
\pgfsetdash{}{0pt}
\pgfsetdash{}{0pt}
\pgfsetbuttcap
\pgfsetmiterjoin
\pgfsetlinewidth{0.100000\du}
\pgfsetbuttcap
\pgfsetmiterjoin
\pgfsetdash{}{0pt}
\definecolor{dialinecolor}{rgb}{0.000000, 0.000000, 0.000000}
\pgfsetstrokecolor{dialinecolor}
\pgfpathellipse{\pgfpoint{19.857500\du}{6.007500\du}}{\pgfpoint{1.487500\du}{0\du}}{\pgfpoint{0\du}{1.487500\du}}
\pgfusepath{stroke}
\pgfsetlinewidth{0.010000\du}
\pgfsetbuttcap
\pgfsetmiterjoin
\pgfsetdash{}{0pt}
\definecolor{dialinecolor}{rgb}{0.000000, 0.000000, 0.000000}
\pgfsetstrokecolor{dialinecolor}
\pgfpathellipse{\pgfpoint{19.857500\du}{6.007500\du}}{\pgfpoint{1.487500\du}{0\du}}{\pgfpoint{0\du}{1.487500\du}}
\pgfusepath{stroke}
\pgfsetlinewidth{0.100000\du}
\pgfsetdash{}{0pt}
\pgfsetdash{}{0pt}
\pgfsetbuttcap
\pgfsetmiterjoin
\pgfsetlinewidth{0.100000\du}
\pgfsetbuttcap
\pgfsetmiterjoin
\pgfsetdash{}{0pt}
\definecolor{dialinecolor}{rgb}{0.000000, 0.000000, 0.000000}
\pgfsetfillcolor{dialinecolor}
\pgfpathellipse{\pgfpoint{19.857500\du}{6.057500\du}}{\pgfpoint{0.737500\du}{0\du}}{\pgfpoint{0\du}{0.737500\du}}
\pgfusepath{fill}
\definecolor{dialinecolor}{rgb}{0.000000, 0.000000, 0.000000}
\pgfsetstrokecolor{dialinecolor}
\pgfpathellipse{\pgfpoint{19.857500\du}{6.057500\du}}{\pgfpoint{0.737500\du}{0\du}}{\pgfpoint{0\du}{0.737500\du}}
\pgfusepath{stroke}
\pgfsetlinewidth{0.010000\du}
\pgfsetbuttcap
\pgfsetmiterjoin
\pgfsetdash{}{0pt}
\definecolor{dialinecolor}{rgb}{0.000000, 0.000000, 0.000000}
\pgfsetstrokecolor{dialinecolor}
\pgfpathellipse{\pgfpoint{19.857500\du}{6.057500\du}}{\pgfpoint{0.737500\du}{0\du}}{\pgfpoint{0\du}{0.737500\du}}
\pgfusepath{stroke}
\pgfsetlinewidth{0.100000\du}
\pgfsetdash{}{0pt}
\pgfsetdash{}{0pt}
\pgfsetbuttcap
{
\definecolor{dialinecolor}{rgb}{0.000000, 0.000000, 0.000000}
\pgfsetfillcolor{dialinecolor}
\definecolor{dialinecolor}{rgb}{0.000000, 0.000000, 0.000000}
\pgfsetstrokecolor{dialinecolor}
\draw (21.345000\du,6.007500\du)--(20.643376\du,6.031084\du);
}
\definecolor{dialinecolor}{rgb}{0.000000, 0.000000, 0.000000}
\pgfsetstrokecolor{dialinecolor}
\node[anchor=west] at (10.450000\du,4.400000\du){1};
\definecolor{dialinecolor}{rgb}{0.000000, 0.000000, 0.000000}
\pgfsetstrokecolor{dialinecolor}
\node[anchor=west] at (17.650000\du,4.500000\du){2};
\definecolor{dialinecolor}{rgb}{0.000000, 0.000000, 0.000000}
\pgfsetstrokecolor{dialinecolor}
\node[anchor=west] at (21.600000\du,4.400000\du){3};
\definecolor{dialinecolor}{rgb}{0.000000, 0.000000, 0.000000}
\pgfsetstrokecolor{dialinecolor}
\node[anchor=west] at (27.850000\du,4.250000\du){4};
\end{tikzpicture}}
\caption[A set of seesaw types..]{A set of seesaw types.}
\label{setseesaw}
\end{center}
\end{figure}
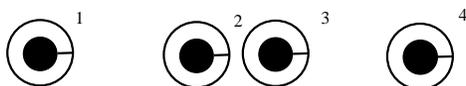
\indent Let us consider a more advanced case. Fig.\ref{cycling} shows a system of four types. For simplicity, there is one agent pro type - agent A for type 1, agent B for type 2 and so on. Agent A can get three different offers. Best matching is the offer of agent D. Worst matching is the offer of agent B. If each agent allures agents with the best and the average matching offer, then each agent will be allured by agents with the average and the worst matching offer. If each agent accept the better one, agent A will cooperate with agent C and agent B will cooperate with agent D. This means that in this cycling offer example, each agent achieves the average satisfaction level.\\
\begin{figure}
\begin{center}
\resizebox{110pt}{!}{
\ifx\du\undefined
  \newlength{\du}
\fi
\setlength{\du}{15\unitlength}
\begin{tikzpicture}
\pgftransformxscale{1.000000}
\pgftransformyscale{-1.000000}
\definecolor{dialinecolor}{rgb}{0.000000, 0.000000, 0.000000}
\pgfsetstrokecolor{dialinecolor}
\definecolor{dialinecolor}{rgb}{1.000000, 1.000000, 1.000000}
\pgfsetfillcolor{dialinecolor}
\pgfsetlinewidth{0.100000\du}
\pgfsetdash{}{0pt}
\pgfsetdash{}{0pt}
\pgfsetbuttcap
\pgfsetmiterjoin
\pgfsetlinewidth{0.100000\du}
\pgfsetbuttcap
\pgfsetmiterjoin
\pgfsetdash{}{0pt}
\definecolor{dialinecolor}{rgb}{0.000000, 0.000000, 0.000000}
\pgfsetstrokecolor{dialinecolor}
\pgfpathellipse{\pgfpoint{27.322500\du}{12.582500\du}}{\pgfpoint{1.487500\du}{0\du}}{\pgfpoint{0\du}{1.487500\du}}
\pgfusepath{stroke}
\pgfsetlinewidth{0.010000\du}
\pgfsetbuttcap
\pgfsetmiterjoin
\pgfsetdash{}{0pt}
\definecolor{dialinecolor}{rgb}{0.000000, 0.000000, 0.000000}
\pgfsetstrokecolor{dialinecolor}
\pgfpathellipse{\pgfpoint{27.322500\du}{12.582500\du}}{\pgfpoint{1.487500\du}{0\du}}{\pgfpoint{0\du}{1.487500\du}}
\pgfusepath{stroke}
\pgfsetlinewidth{0.100000\du}
\pgfsetdash{}{0pt}
\pgfsetdash{}{0pt}
\pgfsetbuttcap
\pgfsetmiterjoin
\pgfsetlinewidth{0.100000\du}
\pgfsetbuttcap
\pgfsetmiterjoin
\pgfsetdash{}{0pt}
\definecolor{dialinecolor}{rgb}{0.000000, 0.000000, 0.000000}
\pgfsetfillcolor{dialinecolor}
\pgfpathellipse{\pgfpoint{19.972500\du}{12.682500\du}}{\pgfpoint{0.737500\du}{0\du}}{\pgfpoint{0\du}{0.737500\du}}
\pgfusepath{fill}
\definecolor{dialinecolor}{rgb}{0.000000, 0.000000, 0.000000}
\pgfsetstrokecolor{dialinecolor}
\pgfpathellipse{\pgfpoint{19.972500\du}{12.682500\du}}{\pgfpoint{0.737500\du}{0\du}}{\pgfpoint{0\du}{0.737500\du}}
\pgfusepath{stroke}
\pgfsetlinewidth{0.010000\du}
\pgfsetbuttcap
\pgfsetmiterjoin
\pgfsetdash{}{0pt}
\definecolor{dialinecolor}{rgb}{0.000000, 0.000000, 0.000000}
\pgfsetstrokecolor{dialinecolor}
\pgfpathellipse{\pgfpoint{19.972500\du}{12.682500\du}}{\pgfpoint{0.737500\du}{0\du}}{\pgfpoint{0\du}{0.737500\du}}
\pgfusepath{stroke}
\pgfsetlinewidth{0.100000\du}
\pgfsetdash{}{0pt}
\pgfsetdash{}{0pt}
\pgfsetbuttcap
{
\definecolor{dialinecolor}{rgb}{0.000000, 0.000000, 0.000000}
\pgfsetfillcolor{dialinecolor}
\definecolor{dialinecolor}{rgb}{0.000000, 0.000000, 0.000000}
\pgfsetstrokecolor{dialinecolor}
\draw (25.835000\du,12.582500\du)--(20.760100\du,12.669100\du);
}
\pgfsetlinewidth{0.100000\du}
\pgfsetdash{}{0pt}
\pgfsetdash{}{0pt}
\pgfsetbuttcap
\pgfsetmiterjoin
\pgfsetlinewidth{0.100000\du}
\pgfsetbuttcap
\pgfsetmiterjoin
\pgfsetdash{}{0pt}
\definecolor{dialinecolor}{rgb}{0.000000, 0.000000, 0.000000}
\pgfsetstrokecolor{dialinecolor}
\pgfpathellipse{\pgfpoint{27.312500\du}{6.197500\du}}{\pgfpoint{1.487500\du}{0\du}}{\pgfpoint{0\du}{1.487500\du}}
\pgfusepath{stroke}
\pgfsetlinewidth{0.010000\du}
\pgfsetbuttcap
\pgfsetmiterjoin
\pgfsetdash{}{0pt}
\definecolor{dialinecolor}{rgb}{0.000000, 0.000000, 0.000000}
\pgfsetstrokecolor{dialinecolor}
\pgfpathellipse{\pgfpoint{27.312500\du}{6.197500\du}}{\pgfpoint{1.487500\du}{0\du}}{\pgfpoint{0\du}{1.487500\du}}
\pgfusepath{stroke}
\pgfsetlinewidth{0.100000\du}
\pgfsetdash{}{0pt}
\pgfsetdash{}{0pt}
\pgfsetbuttcap
\pgfsetmiterjoin
\pgfsetlinewidth{0.100000\du}
\pgfsetbuttcap
\pgfsetmiterjoin
\pgfsetdash{}{0pt}
\definecolor{dialinecolor}{rgb}{0.000000, 0.000000, 0.000000}
\pgfsetfillcolor{dialinecolor}
\pgfpathellipse{\pgfpoint{27.327500\du}{12.682500\du}}{\pgfpoint{0.752500\du}{0\du}}{\pgfpoint{0\du}{0.752500\du}}
\pgfusepath{fill}
\definecolor{dialinecolor}{rgb}{0.000000, 0.000000, 0.000000}
\pgfsetstrokecolor{dialinecolor}
\pgfpathellipse{\pgfpoint{27.327500\du}{12.682500\du}}{\pgfpoint{0.752500\du}{0\du}}{\pgfpoint{0\du}{0.752500\du}}
\pgfusepath{stroke}
\pgfsetlinewidth{0.010000\du}
\pgfsetbuttcap
\pgfsetmiterjoin
\pgfsetdash{}{0pt}
\definecolor{dialinecolor}{rgb}{0.000000, 0.000000, 0.000000}
\pgfsetstrokecolor{dialinecolor}
\pgfpathellipse{\pgfpoint{27.327500\du}{12.682500\du}}{\pgfpoint{0.752500\du}{0\du}}{\pgfpoint{0\du}{0.752500\du}}
\pgfusepath{stroke}
\pgfsetlinewidth{0.100000\du}
\pgfsetdash{}{0pt}
\pgfsetdash{}{0pt}
\pgfsetbuttcap
{
\definecolor{dialinecolor}{rgb}{0.000000, 0.000000, 0.000000}
\pgfsetfillcolor{dialinecolor}
\definecolor{dialinecolor}{rgb}{0.000000, 0.000000, 0.000000}
\pgfsetstrokecolor{dialinecolor}
\draw (27.312500\du,7.685000\du)--(27.325100\du,11.880900\du);
}
\pgfsetlinewidth{0.100000\du}
\pgfsetdash{}{0pt}
\pgfsetdash{}{0pt}
\pgfsetbuttcap
\pgfsetmiterjoin
\pgfsetlinewidth{0.100000\du}
\pgfsetbuttcap
\pgfsetmiterjoin
\pgfsetdash{}{0pt}
\definecolor{dialinecolor}{rgb}{0.000000, 0.000000, 0.000000}
\pgfsetstrokecolor{dialinecolor}
\pgfpathellipse{\pgfpoint{20.172500\du}{6.322500\du}}{\pgfpoint{1.487500\du}{0\du}}{\pgfpoint{0\du}{1.487500\du}}
\pgfusepath{stroke}
\pgfsetlinewidth{0.010000\du}
\pgfsetbuttcap
\pgfsetmiterjoin
\pgfsetdash{}{0pt}
\definecolor{dialinecolor}{rgb}{0.000000, 0.000000, 0.000000}
\pgfsetstrokecolor{dialinecolor}
\pgfpathellipse{\pgfpoint{20.172500\du}{6.322500\du}}{\pgfpoint{1.487500\du}{0\du}}{\pgfpoint{0\du}{1.487500\du}}
\pgfusepath{stroke}
\pgfsetlinewidth{0.100000\du}
\pgfsetdash{}{0pt}
\pgfsetdash{}{0pt}
\pgfsetbuttcap
\pgfsetmiterjoin
\pgfsetlinewidth{0.100000\du}
\pgfsetbuttcap
\pgfsetmiterjoin
\pgfsetdash{}{0pt}
\definecolor{dialinecolor}{rgb}{0.000000, 0.000000, 0.000000}
\pgfsetfillcolor{dialinecolor}
\pgfpathellipse{\pgfpoint{27.272500\du}{6.272500\du}}{\pgfpoint{0.737500\du}{0\du}}{\pgfpoint{0\du}{0.737500\du}}
\pgfusepath{fill}
\definecolor{dialinecolor}{rgb}{0.000000, 0.000000, 0.000000}
\pgfsetstrokecolor{dialinecolor}
\pgfpathellipse{\pgfpoint{27.272500\du}{6.272500\du}}{\pgfpoint{0.737500\du}{0\du}}{\pgfpoint{0\du}{0.737500\du}}
\pgfusepath{stroke}
\pgfsetlinewidth{0.010000\du}
\pgfsetbuttcap
\pgfsetmiterjoin
\pgfsetdash{}{0pt}
\definecolor{dialinecolor}{rgb}{0.000000, 0.000000, 0.000000}
\pgfsetstrokecolor{dialinecolor}
\pgfpathellipse{\pgfpoint{27.272500\du}{6.272500\du}}{\pgfpoint{0.737500\du}{0\du}}{\pgfpoint{0\du}{0.737500\du}}
\pgfusepath{stroke}
\pgfsetlinewidth{0.100000\du}
\pgfsetdash{}{0pt}
\pgfsetdash{}{0pt}
\pgfsetbuttcap
{
\definecolor{dialinecolor}{rgb}{0.000000, 0.000000, 0.000000}
\pgfsetfillcolor{dialinecolor}
\definecolor{dialinecolor}{rgb}{0.000000, 0.000000, 0.000000}
\pgfsetstrokecolor{dialinecolor}
\draw (21.660000\du,6.322500\du)--(26.485300\du,6.279510\du);
}
\pgfsetlinewidth{0.100000\du}
\pgfsetdash{}{0pt}
\pgfsetdash{}{0pt}
\pgfsetbuttcap
\pgfsetmiterjoin
\pgfsetlinewidth{0.100000\du}
\pgfsetbuttcap
\pgfsetmiterjoin
\pgfsetdash{}{0pt}
\definecolor{dialinecolor}{rgb}{0.000000, 0.000000, 0.000000}
\pgfsetstrokecolor{dialinecolor}
\pgfpathellipse{\pgfpoint{20.007500\du}{12.747500\du}}{\pgfpoint{1.497500\du}{0\du}}{\pgfpoint{0\du}{1.497500\du}}
\pgfusepath{stroke}
\pgfsetlinewidth{0.010000\du}
\pgfsetbuttcap
\pgfsetmiterjoin
\pgfsetdash{}{0pt}
\definecolor{dialinecolor}{rgb}{0.000000, 0.000000, 0.000000}
\pgfsetstrokecolor{dialinecolor}
\pgfpathellipse{\pgfpoint{20.007500\du}{12.747500\du}}{\pgfpoint{1.497500\du}{0\du}}{\pgfpoint{0\du}{1.497500\du}}
\pgfusepath{stroke}
\pgfsetlinewidth{0.100000\du}
\pgfsetdash{}{0pt}
\pgfsetdash{}{0pt}
\pgfsetbuttcap
\pgfsetmiterjoin
\pgfsetlinewidth{0.100000\du}
\pgfsetbuttcap
\pgfsetmiterjoin
\pgfsetdash{}{0pt}
\definecolor{dialinecolor}{rgb}{0.000000, 0.000000, 0.000000}
\pgfsetfillcolor{dialinecolor}
\pgfpathellipse{\pgfpoint{20.057500\du}{6.357500\du}}{\pgfpoint{0.737500\du}{0\du}}{\pgfpoint{0\du}{0.737500\du}}
\pgfusepath{fill}
\definecolor{dialinecolor}{rgb}{0.000000, 0.000000, 0.000000}
\pgfsetstrokecolor{dialinecolor}
\pgfpathellipse{\pgfpoint{20.057500\du}{6.357500\du}}{\pgfpoint{0.737500\du}{0\du}}{\pgfpoint{0\du}{0.737500\du}}
\pgfusepath{stroke}
\pgfsetlinewidth{0.010000\du}
\pgfsetbuttcap
\pgfsetmiterjoin
\pgfsetdash{}{0pt}
\definecolor{dialinecolor}{rgb}{0.000000, 0.000000, 0.000000}
\pgfsetstrokecolor{dialinecolor}
\pgfpathellipse{\pgfpoint{20.057500\du}{6.357500\du}}{\pgfpoint{0.737500\du}{0\du}}{\pgfpoint{0\du}{0.737500\du}}
\pgfusepath{stroke}
\pgfsetlinewidth{0.100000\du}
\pgfsetdash{}{0pt}
\pgfsetdash{}{0pt}
\pgfsetbuttcap
{
\definecolor{dialinecolor}{rgb}{0.000000, 0.000000, 0.000000}
\pgfsetfillcolor{dialinecolor}
\definecolor{dialinecolor}{rgb}{0.000000, 0.000000, 0.000000}
\pgfsetstrokecolor{dialinecolor}
\draw (20.019400\du,11.224900\du)--(20.051300\du,7.144550\du);
}
\definecolor{dialinecolor}{rgb}{0.000000, 0.000000, 0.000000}
\pgfsetstrokecolor{dialinecolor}
\node[anchor=west] at (18.650000\du,9.250000\du){1};
\definecolor{dialinecolor}{rgb}{0.000000, 0.000000, 0.000000}
\pgfsetstrokecolor{dialinecolor}
\node[anchor=west] at (23.100000\du,5.150000\du){2};
\definecolor{dialinecolor}{rgb}{0.000000, 0.000000, 0.000000}
\pgfsetstrokecolor{dialinecolor}
\node[anchor=west] at (27.800000\du,9.250000\du){3};
\definecolor{dialinecolor}{rgb}{0.000000, 0.000000, 0.000000}
\pgfsetstrokecolor{dialinecolor}
\node[anchor=west] at (22.950000\du,13.450000\du){4};
\end{tikzpicture}}
\caption[Cycling demand.]{Cycling demand.}
\label{cycling}
\end{center}
\end{figure}
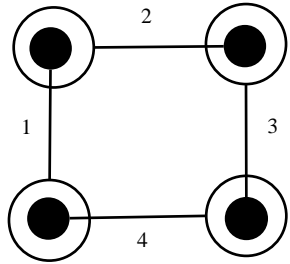
\begin{figure}
\begin{center}
\resizebox{390pt}{!}{\input{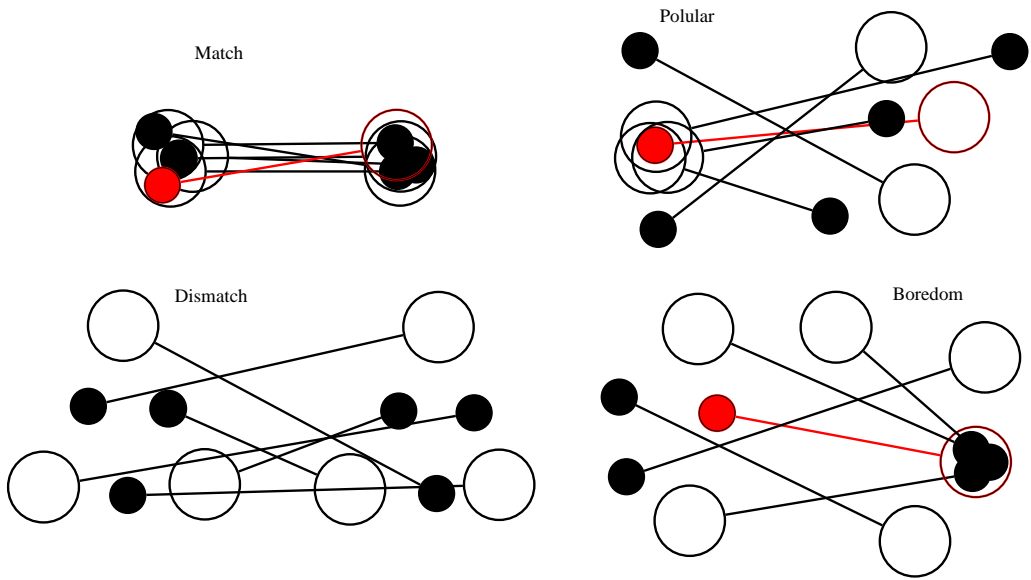}}
\caption[Cases for bipartite agent types.]{Cases for bipartite agent types.}
\label{bipartite}
\end{center}
\end{figure}
\indent The real world examples for the barter double auction are allocations of high level jobs or sexual relationships among humans. The value of a high job offer can not be reduced to the money one can earn there. For a job offer like a postdoc position, one also looks on the climate, country, language, scientific level and so on. The same will be, if one is searching for a marriage partner. Both examples are bipartite (in heterosexual case) systems of agent's type. That means that there are two categories of offers and two categories of demands. If an agent has an offer of the one category, he will have a demand of another category. There can be 4 different cases (Fig.\ref{bipartite}). In case 'match', the agents must only coordinate as in seesaw example. In case 'dismatch', the agents can not achieve any valuable satisfaction level. Most would cooperate at all. In case 'popular', there is an agent (red) which is wooed by agents of with an offer of another category and can choose the best offer. In case 'boredom', there is an agent (red) whose demand is exactly the average offer. 
\section{Conclusions}\label{conclusions}
\indent In this paper, a new game concept is introduced. This concept is the barter double auction. This concept is introduced in a very preliminary way. Nevertheless, it can be used to model non-monetary strategic interactions.
\bibliographystyle{abbrv}
\bibliography{barter}
\end{document}